\documentclass[conference]{IEEEtran}
\IEEEoverridecommandlockouts

\usepackage{cite}
\usepackage{amsmath,amssymb,amsfonts}
\usepackage{graphicx}
\usepackage{textcomp}
\usepackage{xcolor}
\usepackage{amsthm}
\usepackage{bm}
\usepackage{graphicx}
\usepackage{float} 
\usepackage{subfigure}
\usepackage{setspace} 

\usepackage[linesnumbered, ruled]{algorithm2e}
\usepackage{algpseudocode}
\usepackage{ragged2e}

\newtheorem{myDef}{Definition}

\begin{document}

\title{
QoE-Aware Resource Allocation for Semantic Communication Networks
}

\author{\IEEEauthorblockN{Lei Yan\IEEEauthorrefmark{2},
Zhijin Qin
\IEEEauthorrefmark{3},
Rui Zhang\IEEEauthorrefmark{2}, 
Yongzhao Li\IEEEauthorrefmark{2},
and Geoffrey Ye Li\IEEEauthorrefmark{4}}
\IEEEauthorblockA{\IEEEauthorrefmark{2}State Key Laboratory of Integrated Services Networks, Xidian University, Xi'an, China \\
\IEEEauthorrefmark{3}School of Electronic Engineering and Computer Science, Queen Mary University of London, London, UK\\
\IEEEauthorrefmark{4}School of Electrical and Electronic Engineering, Imperial College London, London, UK\\
E-mail: lyan@stu.xidian.edu.cn, z.qin@qmul.ac.uk, \{rz, yzhli\}@xidian.edu.cn, geoffrey.li@imperial.ac.uk
}
}
\maketitle
\vspace{-10pt}
\begin{abstract}
With the aim of accomplishing intelligence tasks, semantic communications transmit task-related information only, yielding significant performance gains over conventional communications. 
To guarantee user requirements for different types of tasks, we perform the semantic-aware resource allocation in a multi-cell multi-task network in this paper.
Specifically, an approximate measure of semantic entropy is first developed to quantify the semantic information for different tasks,
based on which a novel quality-of-experience (QoE) model is proposed. We formulate the QoE-aware semantic resource allocation in terms of the number of transmitted semantic symbols, channel assignment, and power allocation. To solve this problem, we first decouple it into two independent subproblems. The first one is to optimize the number of transmitted semantic symbols with given channel assignment and power allocation, which is solved by the exhaustive searching method. The second one is the channel assignment and power allocation subproblem, which is modeled as a many-to-one matching game and solved by the proposed low-complexity matching algorithm. Simulation results demonstrate the effectiveness and superiority of the proposed method on the overall QoE.
\end{abstract}


\begin{IEEEkeywords}
Quality of experience, resource allocation, semantic communications, semantic-aware networks.
\end{IEEEkeywords}
\section{Introduction}
Different from conventional communications that focus on the engineering problem regardless of semantics and transmission tasks, 
semantic communications are task-oriented where only task-related information is extracted and transmitted \cite{zhijin2021challenges}. To be specific, 
different types of tasks, including single-modal tasks, e.g., text transmission\cite{DeepSC}
and image retrieval\cite{SingleModalImage}, and multimodal tasks, e.g., visual question answering (VQA)\cite{DeepSC-VQA}, require different semantic transceivers to ensure the superior performance in transmission reliability and efficiency. Therefore, the conventional resource allocation model is not applicable and the resource allocation for such a semantic-aware network need to be revisited.





Some researchers have made several preliminary studies on this issue. 
In \cite{RAPerformanceOptimization} and \cite{SemanticCompressionPC}, the importance weights of the extracted semantic features are taken into account. Particularly, in addition to the conventional optimization variables, the partial semantic information to be transmitted was optimized in \cite{RAPerformanceOptimization} for the text transmission task. The compression ratio of semantic features has been investigated in \cite{SemanticCompressionPC} for the image classification task. However, both works aimed to maximize the task performance, without evaluating the semantic communication efficiency. 
Although Xia \textit{et al.} \cite{infocom} maximized the system throughput in message for text transmission, considering the dynamic background knowledge matching condition, they optimized the resource allocation at the message level rather than the semantic level.


To investigate the resource allocation at the semantic level, 
our prior work \cite{OurWork} has defined the semantic transmission rate (S-R) and the semantic spectral efficiency (S-SE), and formulated a problem to maximize the overall S-SE. Nevertheless, since the parameters related to the semantic information quantification are constants for a single task and are ignored to obtain the solutions, the work can only be used to allocate resources for the users with the same task.
To cope with the coexistence of multiple tasks, we conduct further study on the semantic-aware resource allocation in this paper.

Particularly, since the performance of semantic communications depends on whether the user can complete the task, focusing more on the subjective user experience, semantic communications inherently have the advantage of enhancing quality-of-experience (QoE) \cite{shiQoE}, which considers both objective quality-of-service (QoS) and subjective user experience\cite{QoESwap}. In this regard, we study the semantic-aware resource allocation based on the QoE metric in a multi-cell multi-task network. The main contributions of this paper are summarized as follows: 
\vspace{-1pt}
\begin{itemize}
    \item A QoE-aware resource allocation problem is formulated for semantic communication networks.
    Specifically, based on the developed approximate semantic entropy, a novel QoE model is used to formulate the optimization problem in terms of the number of transmitted semantic symbols, channel assignment, and power allocation.
    \item The formulated problem is decoupled into two subproblems, which are solved by the exhaustive searching method and a low-complexity matching algorithm, respectively.
    
    
    \item Simulation results
    verify the superiority of the proposed QoE-aware semantic resource allocation method in terms of the overall QoE against the baselines.
\end{itemize}

\vspace{-1pt}

\textit{Notation:} Bold-font variables represent matrices and vectors. Calligraphic-font variables represent sets. The superscript, $\mathbf{H}^{\rm{H}}$, represents the conjugate transpose of $\mathbf{H}$. $\mathbb{E}(x)$ is the expectation of $x$. In addition, $x\sim U(a,b)$ means that $x$ follows a uniform distribution over the interval $[a,b]$, and $x\sim N(\mu,\sigma^2)$ means that $x$ follows a normal distribution with mean $\mu$ and covariance $\sigma^2$.


\section{System Model}
In this article, we consider an uplink cellular network with $B$ coordinated cells, where each cell has a base station (BS) with $N_{\rm{r}}$ receiving antennas and a set of single antenna users. Let $\mathcal{U}^b$ denote the user set in the $b$-th cell. We assume that $B$ BSs form a cooperation cluster, which can exchange channel state information (CSI) and scheduling information with a central controller through high-speed low-delay fiber backhaul links \cite{LuCooperation}. In addition, we consider the coexistence of different intelligence tasks in the network, including single-modal tasks and multimodal tasks. Task-specific semantic communication transmitters and receivers are equipped at users and BSs, respectively.
The transmission model and semantic communication models are introduced below.

\subsection{Transmission Model}
In order to fully exploit spectrum resources, we assume that the $B$ coordinated cells share the same channel set denoted by  $\mathcal{M}$, and each user can occupy at most one channel with bandwidth $W$. In each cell, the channels are orthogonally allocated to users to eliminate the intra-cell interference. With the maximal ratio combining (MRC) detection at the BS, the receiving signal-to-interference-plus-noise ratio (SINR) of user $u, u\in \mathcal{U}^b$ at the $b$-th BS can be expressed as
\begin{equation}
    \gamma_u^b=\sum\limits_{m \in \mathcal{M}} \frac{\alpha_{u,m} p_{u} |{\bf{W}}_{u,m}^{b} {\bf{H}}_{u,m}^{b} |^2}{\|{\bf{W}}_{u,m}^{b} \|^2 \sigma^2+I_{u,m}^{b}},
\end{equation}
where $\alpha_{u,m}\in \{0,1\}$, $\alpha_{u,m}=1$ if channel $m \in \mathcal{M}$ is allocated to user $u$ and $\alpha_{u,m}=0$ otherwise, $p_{u}$ is the transmit power of user $u$, and ${\bf{H}}_{u,m}^{b}$ is the $N_{\rm{r}} \times 1$ channel matrix from user $u$ to the $b$-th BS over channel $m$. Accordingly, the MRC detection matrix can be obtained by ${\bf{W}}_{u,m}^{b}=({\bf{H}}_{u,m}^{b})^{\rm{H}}$. In (1), $\sigma^2$ is the noise power spectral density and $I_{u,m}^{b}$ is the interference experienced by user $u$ over channel $m$ from the adjacent cells, that is
\begin{equation}
    {I}_{u,m}^{b} = \sum\limits_{b'\in\mathcal{B},b'\ne b} \sum\limits_{u'\in \mathcal{U}^{b'}} \alpha_{u',m} p_{u'} |{\bf{W}}_{u,m}^{b} {\bf{H}}_{u',m}^{b} |^2.
\end{equation}

\subsection{Semantic Communication Models}
We focus on two types of intelligence tasks in this paper, including a single-modal task and a bimodal task. However, the proposed algorithm can be extended to the case of multiple multimodal intelligence tasks easily. 
Assume that $N_{\rm{Si}}$ single-modal users and $N_{\rm{Bi}}$ bimodal user pairs are randomly deployed in the considered network, and the $b$-th cell has $N_{\rm{Si}}^{b}$ single-modal users and $N_{\rm{Bi}}^b$ bimodal user pairs.
For ease of clarification, a single-modal user or a bimodal user pair is regarded as a user group. The index set of all user groups in the $b$-th cell is given by
\begin{equation}
\begin{split}
    {\mathcal{Q}}^b&=\{1,2,\dots,q,\dots,|{\mathcal{Q}}^b|\}\\
    &=\{1,2,\dots,N_{\rm{Bi}}^b,N_{\rm{Bi}}^b+1,\dots,N_{\rm{Bi}}^b+N_{\rm{Si}}^{b}\},
\end{split}
\end{equation}
where $|{\mathcal{Q}}^b|=N_{\rm{Bi}}^b+N_{\rm{Si}}^{b}$. Accordingly, the $q$-th user group contains a bimodal user pair when $q \le N_{\rm{Bi}}^b$, and it contains a single-modal user when $q > N_{\rm{Bi}}^b$. Besides, we denote the set of all user groups in the $b$-th cell by $\mathcal{G}^b=\{\mathcal{G}_q^b\}_{q\in \mathcal{Q}^b}$ where $\mathcal{G}_q^b \subseteq \mathcal{U}^b$ represents the user set of the $q$-th group.

For the single-modal task, we take text transmission task as an example. Specifically, DeepSC \cite{DeepSC} is adopted, where the sentence generated by the user is first mapped to semantic symbols that can be directly transmitted over the physical channel by the DeepSC transmitter, and then recovered by the DeepSC receiver at the BS. Moreover, semantic similarity is used to evaluate the performance, which is a function of the number of transmitted semantic symbols and SINR \cite{OurWork}. Thus, the semantic similarity of the user group $\mathcal{G}_q^b, q> N_{\rm{Bi}}^b$ can be expressed as $\xi_{q}^b=f_{\rm{Si}}(k_u,\gamma_{u}^b), u\in \mathcal{G}_q^b$ where $k_u$ is the average number of transmitted semantic symbols of user $u$.

For the bimodal task, we take VQA task as an example and adopt the developed DeepSC-VQA model \cite{DeepSC-VQA}. This task involves two users for text and image transmission, respectively. The two users first extract the semantic symbols from the text and image information through the DeepSC-VQA transmitter, respectively, and then send them to the BS. The received semantic symbols of text and image will be fused by the DeepSC-VQA receiver at the BS to predict the answer. As the two users jointly decide the task performance, the answer accuracy of this task could be modeled as a function with respect to the numbers of transmitted semantic symbols and the SINR of the two users, i.e., $\xi_{q}^b=f_{\rm{Bi}}(k_{u_{\rm{t}}}, k_{u_{\rm{i}}},\gamma_{u_{\rm{t}}}^b,\gamma_{u_{\rm{i}}}^b), u_{\rm{t}},u_{\rm{i}}\in \mathcal{G}_q^b$, where $u_{\rm{t}}$ and $u_{\rm{i}}$ represent the user for text transmission and the one for image transmission, respectively.



\section{Problem Formulation}
In this section, we first develop a method to obtain an approximate measure of semantic entropy, based on which a novel QoE model for semantic communications is proposed. Then, the semantic resource allocation is formulated as a QoE maximization problem.
\subsection{Approximate Measure of Semantic Entropy} 
Information entropy measures the information based on the statistical characteristic of source symbols, while semantic entropy directly quantifies the semantic information of the source and plays a significant role in evaluating semantic communication efficiency and resource allocation. However, a common semantic information metric is still missing, and most existing metrics are non-calculable \cite{zhijin2021challenges}. 

Since the semantic information carried by a source depends on the specific task, 
semantic entropy should be a measure with respect to the source and the task. 
Following Chattopadhyay \textit{et al.}~\cite{chattopadhyay2020quantifying}, we define the semantic entropy as following:
\begin{myDef}
Given semantic source $\mathcal{X}$, semantic entropy is defined as the minimum number of semantic symbols about the data $X \in \mathcal{X}$ that are sufficient to predict task $Y$, i.e.,
\vspace{-5pt}
\begin{equation}
\begin{split}
    &H(X;Y)\buildrel \Delta \over =\mathop {\min }\limits_E \mathbb{E}\left(\left| Code^E(X) \right|\right) \\
    &{\rm{s.t.}}\quad P(Y|Code^E(X))=P(Y|X),
\end{split}
\vspace{-5pt}
\end{equation}
where $Code^E(x)$ denotes the semantic symbol vector extracted from $X$ with the semantic encoder $E$, and $P(Y|X)$ is the conditional probability of~ $Y$ given $X$.
\end{myDef}
From Definition 1, the semantic entropy of $X$ given $Y$ is actually defined as an expected value over the whole data set $\mathcal{X}$, i.e., the semantic entropy is a constant for the same task and dataset, which shares a similar philosophy as \cite{OurWork}. 
However, it is intractable to find an optimal $E$ to derive the semantic entropy. To obtain a measure that is both meaningful and manipulable, we utilize a well-designed deep learning (DL) model as the semantic encoder to obtain the approximate semantic entropy for a task, which can be expressed as 
\begin{equation}
\begin{split}
    &{\tilde H}(X;Y)\buildrel \Delta \over =\mathop {\min } \mathbb{E}\left(\left| Code^{E_{\rm{DL}}}(X) \right|\right) \\
    {\rm{s.t.}}&\quad P(Y|X)-P(Y|Code^{E_{\rm{DL}}}(X))<\varepsilon,
\end{split}
\vspace{-3pt}
\end{equation}
where the constraint indicates that the gap between $P(Y|X)$ and $P(Y|Code^{E_{\rm{DL}}}(X))$ can not exceed $\varepsilon$.

According to the aforementioned method, the approximate semantic entropy of the considered tasks can be derived based the corresponding DL models. In particular, we first remove the channel models from DeepSC and DeepSC-VQA, then train them under different settings of the number of semantic symbols, and finally find the minimum number of semantic symbols that can guarantee a performance very close to the upper bound. Furthermore, we define the unit of the approximate semantic entropy as \textit{sut} as in \cite{OurWork}.
\subsection{QoE Model for Semantic Communications}
We formulate the QoE model based on two objective metrics, \textit{semantic accuracy} and \textit{semantic rate}. The former, the accuracy of message transmission, corresponds to the semantic similarity in the considered single-modal task and the answer accuracy in the considered bimodal task. The latter is defined as the amount of semantic information emitted to the transmission medium per second, measured in \textit{suts/s}. The semantic rate of the single-modal user $u\in \mathcal{G}_q^b, q> N_{\rm{Bi}}^b$ is given as
\begin{equation}
    \varphi_u=\frac{\tilde H_{\rm{Si}}}{k_u/W},
\end{equation}
where $\tilde H_{\rm{Si}}$ represents the DeepSC\cite{DeepSC} based approximate semantic entropy. 
The semantic rate of bimodal users $u_{\rm{t}}$ and $u_{\rm{i}}$, $u_{\rm{t}},u_{\rm{i}}\in \mathcal{G}_q^b,q \le N_{\rm{Bi}}^b$ are expressed as
\begin{equation}
    \varphi_{u_{\rm{t}}}=\frac{\tilde H_{\rm{Bi,t}}}{k_{u_{\rm{t}}}/W}, \quad {\rm{and}}\ \varphi_{u_{\rm{i}}}=\frac{\tilde H_{\rm{Bi,i}}}{k_{u_{\rm{i}}}/W},
\end{equation}
respectively, where $\tilde H_{\rm{Bi,t}}$ and $\tilde H_{\rm{Bi,i}}$ represent the DeepSC-VQA\cite{DeepSC-VQA} based approximate semantic entropy for text transmission user and image transmission user, respectively.

Note that the semantic rate is significantly different from the S-R that is defined as the amount of \textit{successfully} delivered semantic information per second in \cite{OurWork}. With $\Gamma_u$ denoting the S-R, we have $\Gamma_u=\varphi_u\xi_q$, i.e., the difference between them lies in that whether the semantic accuracy is considered. The S-R couples semantic accuracy and semantic rate. However, from the perspective of users, accuracy and efficiency of message transmission are different, and users may have their own preferences on them depending on the applications. For example, some users prefer high accuracy but are delay-tolerated, while others may desire higher semantic rate but do not need a very high accuracy. In order to reflect the QoE requirements of users more properly, we consider the two parameters to formulate the QoE model as
\begin{equation}
\begin{split}
    QoE_q^b&=\sum\limits_{u\in{\mathcal{G}_q^b}}{{w_u}G_u^{\rm{R}}+(1-w_u)G_u^{\rm{A}}}\\
    &=\sum\limits_{u\in{\mathcal{G}_q^b}}{\frac{{w_u}}{1+e^{\beta_u(\varphi_u^{\rm{req}}-\varphi_u)}}+\frac{(1-w_u)}{1+e^{\lambda_u(\xi_u^{\rm{req}}-\xi_q^b)}}},
\end{split}
\vspace{-3pt}
\end{equation}
where $QoE_q^b$ is the QoE of the $q$-th user group in the $b$-th cell, $w_u$ and $(1-w_u)$ are the weights of semantic rate and semantic accuracy at user $u$ respectively, $G_u^{\rm{R}}$ and $G_u^{\rm{A}}$ are the scores of semantic rate and semantic accuracy at user $u$ respectively, and $\beta_u$ and $\lambda_u$ represent the growth rates of $G_u^{\rm{R}}$ and $G_u^{\rm{A}}$ respectively. Additionally, $\varphi_u^{\rm{req}}$ and $\xi_u^{\rm{req}}$ represent the minimum semantic rate and semantic accuracy to acquire the 50\% of the scores, respectively. Due to the simplicity and generality, we choose the logistic function to model the correlation between QoE and QoS metrics while other QoE function types are applicable as well for the proposed method, such as MOS-based function or exponential function. Here, $QoE_q^b$, $G_u^{\rm{R}}$, and $G_u^{\rm{A}}$ are between 0 and 1.




\subsection{Problem Formulation}
In this part, we formulate the semantic-aware resource allocation with the goal of maximizing the overall QoE of all users in terms of channel assignment, power allocation, and the number of transmitted semantic symbols. The optimization problem can be expressed as follows:
\begin{align}
    {\rm{\mathbf{(P0)}}}\quad & \mathop{\max}\limits_{\{\{\alpha_{u,m}\},\{p_{u}\},\{k_{u}\}\}}\  {\sum\limits_{b \in \mathcal{B}}{\sum\limits_{q\in{\mathcal{Q}^b}}{QoE_q^b}}}\label{AA}\\
    {\rm{ s.t.}}\ \ &{\rm{ C_1 }}:\  \alpha _{u,m} \in \{0,1\},\  \forall u \in  {\mathcal{U}}^b,\ \!\forall m \!\in\! \mathcal{M},\ \forall b \!\in\! \mathcal{B} \tag{\ref{AA}{a}}, \label{AAa}\\
    &{\rm{ C_2 }}:\  \sum\limits_{u \in  {\mathcal{U}}^b} {{\alpha _{u,m}} \le 1,\ \forall m \in \mathcal{M}},\ \forall b \!\in\! \mathcal{B}\tag{\ref{AA}{b}}, \label{AAb}\\
    &{\rm{    C_3   }}:\  \sum\limits_{m\in \mathcal{M}}{{\alpha _{u,m}} \le 1,\ \forall u \in {\mathcal{U}}^b},\ \forall b \!\in\! \mathcal{B} \tag{\ref{AA}{c}}, \label{AAc}\\
    &{\rm{     C_4  }}:\  \sum\limits_{u\in \mathcal{G}_q^b,q \le N_{\rm{Bi}}^b}{{\alpha _{u,m}} \in \{0,2\},\ \forall m \in {\mathcal{M}}},\ \forall b \!\in\! \mathcal{B} \tag{\ref{AA}{d}}, \label{AAd}\\
    &{\rm{     C_5  }}:\  {k_u} \in {\mathcal{K}}_u,\ \forall u \in {\mathcal{U}}^b,\ \forall b \!\in\! \mathcal{B} \tag{\ref{AA}{e}}, \label{AAe}\\
    &{\rm{     C_6  }}:\   0 \le p_{u} \le P_{\rm{max}},\ \forall u \in {\mathcal{U}}^b,\ \forall b \!\in\! \mathcal{B} \tag{\ref{AA}{f}}, \label{AAf}\\
    &{\rm{C_7}}:\  G_{u}^{\rm{R}}, G_{u}^{\rm{A}} \ge G_{\rm{th}},\ \forall u \in {\mathcal{U}}^b,\ \forall b \!\in\! \mathcal{B} \tag{\ref{AA}{g}} \label{AAg}
    \text{,}
\end{align}
where $\rm{C_1}$ constrains the range of $\alpha_{u,m}$, $\rm{C_2}$ ensures the orthogonal channels for each user in a cell, $\rm{C_3}$ restricts each user to occupy at most one channel, $\rm{C_4}$ ensures that a bimodal user pair will be allocated no channel or two channels as only one channel assigned to one of them will lead to a failure, $\rm{C_5}$ specifies the range of the number of transmitted semantic symbols for each user, and  $\mathcal{K}_u=\mathcal{K}_{\rm{Si}}, \mathcal{K}_{\rm{Bi,t}}, {\rm{and}} \  \mathcal{K}_{\rm{Bi,i}}$ when $u$ is the user with the single-modal task, the user with the bimodal task for text transmission, and the user with the bimodal task for image transmission, respectively, $\rm{C_6}$ constrains the range of transmit power, and $\rm{C_7}$ limits the minimum required scores of semantic rate and semantic accuracy.

\section{A Matching Theory Based Solution}

In this section, we first decouple ${\rm{\mathbf{(P0)}}}$ into two independent subproblems, and then solve them respectively to obtain a suboptimal solution.

\subsection{Problem Decoupling}
By observing ${\rm{\mathbf{(P0)}}}$, the optimization variables, $\{\alpha_{u,m}\}$ and $\{p_u\}$, jointly decide the SINR $\gamma_u^b$. Additionally, the objective function can be expressed as a function with respect to $\gamma_u^b$ and $k_u$. Therefore, given $\gamma_u^b$, the optimal $\{k_u\}$ for each $\mathcal{G}_q^b$ can be obtained by solving the following optimization problem:
\begin{align}
    {\rm{\mathbf{(P1)}}}\quad  &\mathop{\max}\limits_{\{k_{u}\},u\in \mathcal{G}_q^b}\  {{{QoE_q^b}}}\label{YY}\\
    &{\rm{ s.t.}}\ \  {\rm{C_5}}\ {\rm{and}}\ {\rm{C_7}} \nonumber
    \text{.}
\end{align}
Then the channel assignment and power allocation problem can be written as
\begin{align}
    {\rm{\mathbf{(P2)}}}\quad & \mathop{\max}\limits_{\{\{\alpha_{u,m}\},\{p_{u}\}\}}\  {\sum\limits_{b \in \mathcal{B}}{\sum\limits_{q\in{\mathcal{Q}^b}}{QoE_q^{b}}}}\label{YY}\\
    &{\rm{ s.t.}}\ \  {\rm{C_1}},{\rm{C_2}},{\rm{C_3}},{\rm{C_4}},\ {\rm{and}}\ {\rm{C_6}} \nonumber
    \text{.}
\end{align}
Therefore, for each option of $\{\{\alpha_{u,m}\},\{p_{u}\}\}$, the SINR of each user in the network can be calculated using equation (1). Then the optimal $k_u$ and the corresponding maximum ${QoE_q^{b}}$ can be obtained by solving ${\rm{\mathbf{(P1)}}}$. Since $\xi_q^b$ can only be obtained by the look-up table method, the exhaustive searching method is utilized to solve ${\rm{\mathbf{(P1)}}}$ here. Note that the complexity is acceptable due to the limitation of $|{{\mathcal{K}}_u}|$. Finally, with the obtained maximum $QoE_q^{b}$, we can solve ${\rm{\mathbf{(P2)}}}$ to get the solutions, which will be detailed in the next subsection.
\subsection{Channel Assignment and Power Allocation Subproblem}
To cope with the tight coupling among users in multiple cells and those in a bimodal user pair, we construct a matching game to model ${\rm{\mathbf{(P2)}}}$ and propose a low-complexity matching algorithm to obtain the stable matching in this part.

Problem ${\rm{\mathbf{(P2)}}}$ is a three-sided many-to-one matching game among users, channels, and power levels, where each user can select at most one channel and one  power level while each channel or power level can serve multiple users. For the convenience of algorithm implementation, we put all combinations of channels and power levels together to form a resource set $\mathcal{T}=\{(m,p),\forall m \in \mathcal{M}, \forall p\in \mathcal{P}\}$ where $\mathcal{P}$ denotes the power set. Then the three-sided many-to-one matching can be converted to a two-sided matching. In addition, since the QoE of each user depends not only on the opposite partners to be matched, but also on the users of other cells sharing the same channel. More specifically, this is a matching problem with externalities. The idea of swap matching \cite{SwapMatching} can be used for reference to obtain the solution.

Considering two cases may exist in each cell, i.e., $|\mathcal{U}^b|\le|\mathcal{M}|$ and $|\mathcal{U}^b|>|\mathcal{M}|$, the markets of different sides will be focused on to maximize the overall QoE. In particular, the proposed algorithm aims to keep the utility of users increasing by the swap operation to achieve the stable matching for the first case, while focuses on the utility of channels for the second case. 

\textbf{Case 1: $|\mathcal{U}^b|\le|\mathcal{M}|$.} Firstly, we add $|\mathcal{M}|-|\mathcal{U}^b|$ virtual single-modal users to enable every channel to match with a user, making the swap operation easier. Then, the user group set in the $b$-th cell can be updated as ${\mathcal{G}_{\rm{A}}^b}=\mathcal{G}^b\cup \mathcal{U}_{\rm{0}}^b$, where $\mathcal{U}_{\rm{0}}^b$ is the established virtual single-modal user set with $|\mathcal{U}_{\rm{0}}^b|=|\mathcal{M}|-|\mathcal{U}^b|$. Correspondingly, we denote the updated index set of ${\mathcal{G}_{\rm{A}}^b}$ as ${\mathcal{Q}_{\rm{A}}^b}$. 

As the users belonging to a user group jointly decide the task performance, a bimodal user pair (\textit{rather than an individual bimodal user}) or a single-modal user is regarded as a selfish and rational player to make decisions, i.e., there are totally $|{\mathcal{G}_{\rm{A}}^b}|$ players in the $b$-th cell and each of them is denoted by $\mathcal{G}_q^b \in \mathcal{G}_{\rm{A}}^b$. Clearly, the resource sets of the players with the single-modal task and those with the bimodal task are different, which are $\mathcal{T}_{\rm{Si}}=\mathcal{T}$ and $\mathcal{T}_{\rm{Bi}}=\{(m,m',p,p'),\forall m,m' \in \mathcal{M}, m\ne m', \forall p,p' \in \mathcal{P}\}$, respectively. Thus, by denoting $\mathcal{T}_q^b$ as the resource set of the player $\mathcal{G}_q^b$, we have $\mathcal{T}_q^b=\mathcal{T}_{\rm{Bi}}$ when $q\le N_{\rm{Bi}}^b$ and $\mathcal{T}_q^b=\mathcal{T}_{\rm{Si}}$, otherwise. 

Then, we can define a matching $\Phi$ as a function from set $\mathcal{G}_{\rm{A}} \cup \mathcal{T}_{\rm{A}}$ mapping into set $\mathcal{G}_{\rm{A}} \cup \mathcal{T}_{\rm{A}}$, where $\mathcal{G}_{\rm{A}}=\bigcup_{b\in{\mathcal{B}}}\mathcal{G}_{\rm{A}}^b$ and $\mathcal{T}_{\rm{A}}={\mathcal{T}_{\rm{Si}}} \cup {\mathcal{T}_{\rm{Bi}}}$. For a matching pair $(\mathcal{G}_q^b,t), t \in \mathcal{T}_{\rm{A}}$, we have $\Phi(\mathcal{G}_q^b)=t$ and $\Phi(t)=\mathcal{G}_q^b$. The utility function of player $\mathcal{G}_q^b$ under matching $\Phi$ is defined as
\begin{equation}
    U_q^b(\Phi)=\left\{
    \begin{split}
        &QoE_q^b,\quad {\rm{if}}\ q \in {\mathcal{Q}^b};\\
        &0, \quad \quad\quad otherwise.
    \end{split}
    \right.
\end{equation}
Based on this, whether the considered matching is stable can be judged from the following definition.
\begin{myDef}
A matching $\Phi$ is stable if and only if, for each player $\mathcal{G}_q^b \in \mathcal{G}_{\rm{A}}^b, \forall b \in \mathcal{B}$ with $\Phi(\mathcal{G}_q^b)=t$, there does not exist a blocking matching $\Phi_{q}^{\rm{B}}$ such that, $\exists t'\in \mathcal{T}_q^b$:
\begin{enumerate}
    \item $\forall i \in \mathcal{G}_{(t)}\cup \mathcal{G}_{(t')}, U_i^b(\Phi_q^{(t,t')}) \ge U_i^b(\Phi_q^{\rm{B}})$ and
    \item $\exists i \in \mathcal{G}_{(t)}\cup \mathcal{G}_{(t')}, U_i^b(\Phi_q^{(t,t')}) > U_i^b(\Phi_q^{\rm{B}})$,
\end{enumerate}
where $\mathcal{G}_{(t)}$ is formed by all players who have at least one same channel as those included in $t$, and $\Phi_q^{(t,t')}$ represents the swap matching, where the player $\mathcal{G}_q^b$ swap $t$ with $t'$, the players who are in both $\mathcal{G}_{(t')}$ and the $b$-th cell swap their channels accordingly, and the remaining players keep unchanged.
\end{myDef}

\textbf{Case 2: $|\mathcal{U}^b|>|\mathcal{M}|$.} Evidently, if we follow the solution for \textbf{Case 1}, no channel will be tentatively matched with $|\mathcal{U}^b|-|\mathcal{M}|$ of all users even if the overall QoE will increase. To deal with this problem, we focus on the market of channels to achieve the stable matching. 
However, when a channel is matched with a user with the bimodal task, its utility depends not only on which user it matches but on which channel the other user in the same group matches. Hence, we propose the user groups to perform the swap operation but the channels to make the swap decision based on their utility changes.

Similar to \textbf{Case 1}, we first add $|\mathcal{U}^b|-|\mathcal{M}|$ virtual channels to enable the swap operation at each user. Then, the channel set in the $b$-th cell can be updated as ${\mathcal{M}_{\rm{A}}^b}=\mathcal{M}\cup \mathcal{M}_{\rm{0}}^b$, where $\mathcal{M}_{\rm{0}}^b$ is the established virtual channel set with $|\mathcal{M}_{\rm{0}}^b|=|\mathcal{U}^b|-|\mathcal{M}|$. Then, the resource sets can be reformed in the similar manner but based on ${\mathcal{M}_{\rm{A}}^b}$. For simplicity, we use the same notation $\mathcal{T}_q^b$ to denote the resource set of each player $\mathcal{G}_q^b \in \mathcal{G}^b$. Furthermore, the utility function of channel $m\in {\mathcal{M}_{\rm{A}}^b}$ in the $b$-the cell under a matching $\Phi$ is defined as
\begin{equation}
    U_m^b(\Phi)=\left\{
    \begin{split}
        &0, \quad \quad\quad\quad\  {\rm{if}}\  m \in \mathcal{M}_{\rm{0}}^b;\\
        &U_{\mu(m)}^b(\Phi)/2, {\rm{if}}\ m \in \mathcal{M}^b, \mu(m)\le N_{\rm{Bi}}^b;\\
        &U_{\mu(m)}^b(\Phi),\quad{\rm{if}}\ m \in \mathcal{M}^b,  \mu(m)> N_{\rm{Bi}}^b.\\
    \end{split}
    \right.
\end{equation}
where $\mu(m)\in\mathcal{Q}^b$ represents the index of the player that matches channel $m$. When the player is a bimodal user pair, $U_m^b(\Phi)$ is set as the half of the utility of the player, and when the player is a single-modal user, $U_m^b(\Phi)$ is equal to the utility of the player. Then, we give the following definition.
\begin{myDef}
A matching $\Phi$ is stable if and only if, for each channel $m \in {\mathcal{M}_{\rm{A}}^b}$ with $\mu(m)\in\mathcal{Q}^b$, there does not exist a blocking matching $\Phi_{m}^{\rm{B}}$ such that, $\exists t'\in \mathcal{T}_{\mu(m)}^b$:
\begin{enumerate}
    \item $\forall i \in \mathcal{M}_{(\mathcal{G})}, U_i^b(\Phi_{\mu(m)}^{(t,t')}) \ge U_i^b(\Phi_m^{\rm{B}})$ and
    \item $\exists i \in \mathcal{M}_{(\mathcal{G})}, U_i^b(\Phi_{\mu(m)}^{(t,t')}) > U_i^b(\Phi_m^{\rm{B}})$,
\end{enumerate}
where $\mathcal{M}_{(\mathcal{G})}$ is formed by all channels of all players in $\mathcal{G}, \mathcal{G}=\mathcal{G}_{(t)}\cup \mathcal{G}_{(t')}$.
\end{myDef}

Based on the two definitions, we develop an efficient matching algorithm to obtain the stable matching, as shown in Algorithm 1. The proposed algorithm starts by a random initial matching. Then the users will search their resource set to find the blocking matching and thus update the current matching. Once no blocking matching is found, the stable matching $\Phi$ will be the output.

\setlength{\textfloatsep}{3pt}
\begin{algorithm}
\SetAlgoVlined 
\caption{ Proposed Matching Algorithm for Channel Assignment and Power Allocation}
\label{alg1}
\KwIn{$\mathcal{B},\ {\mathcal{U}^b}, \ {\mathcal{M}_{\rm{A}}^b},\ {\mathcal{G}_{\rm{A}}^b}, \ \mathcal{T}_q^b,\ \forall q\in \mathcal{Q}_{\rm{A}}^b, \ \forall b\in\mathcal{B}$.}
{\bf Initialization:} For each cell $b\in \mathcal{B}$, initial the matching channels and power levels of all users with a permutation of all channels and the minimum power level, respectively. Denote the current matching as $\Phi$.\\
\Repeat{No blocking matching is found}{
\For{all $b\in \mathcal{B}$}
{
\For{all $\mathcal{G}_q^b\in \mathcal{G}_{\rm{A}}^b$}
{
$t=\Phi(\mathcal{G}_q^b)$;\\
\For{all $t'\in \mathcal{T}_q^b,\ t'\ne t$}{
\eIf{$|\mathcal{U}^b|\le|\mathcal{M}|$}
{If $\Phi$ is a blocking matching according to Definition 2, update the matching as $\Phi=\Phi_q^{(t,t')}$; otherwise, keep the current matching state.}
{If $\Phi$ is a blocking matching according to Definition 3, update the matching as $\Phi=\Phi_{\mu(m)}^{(t,t')}$ where $m$ is the channel in $t$; otherwise, keep the current matching state.
}
}}
}
}
\KwOut{the stable matching $\Phi$.}
\end{algorithm}

\textit{Remark 1:} The worst-case complexity of Algorithm 1 is as $\mathcal{O}(V \times |\mathcal{B}|\times((|\mathcal{G}_{\rm{A}}^b|-N_{\rm{Bi}}^b)\times|\mathcal{M}_{\rm{A}}^b|\times|\mathcal{P}|+N_{\rm{Bi}}^b \times \tbinom{2}{|\mathcal{M}_{\rm{A}}^b|}\times|\mathcal{P}|^2))$ where $V$ is the number of iterations. However, the complexity of exhaustive searching method here is as $\mathcal{O}(({\frac{{\
\rm{max}}(|\mathcal{U}^b|,M)!}{{\
\rm{min}}(|\mathcal{U}^b|,M)!}\times |\mathcal{P}|^{|\mathcal{U}^b|}})^{|\mathcal{B}|})$.

\textit{Remark 2:} Since the utility of user groups or channels is bounded by 1 and will increase monotonically by the swap operation, Algorithm 1 will terminate to a final matching after a finite number of iterations.
\section{Simulation results}
In this section, we present simulation results to demonstrate the performance of the proposed QoE-aware resource allocation method in a multi-cell multi-task network. 

We consider three cells in the simulation, where users are randomly deployed. The radius of each cell is set as $500\ {\rm{m}}$. Each BS is equipped with two receiving antennas and each user is with single antenna. For the channel model, both large-scale fading, including pathloss and shadowing, and small-scale Rayleigh fading are considered. The adopted pathloss model is $128.1+37.6\rm{lg[d(km)]\  dB}$ and the shadowing factor is set as $6\ {\rm{dB}}$. The bandwith of each channel is $W=180\ \rm{kHz}$. The noise power spectral density is $\sigma^2=-174\ {\rm{dBm/Hz}}$. 
In the simulation, we consider the available power levels as $\mathcal{P}=\{-10, -5, 0, 5, 10, 15, 20\}\ {\rm{dBm}}$. 
Meanwhile, each user generates its QoE-related parameters as $w_u\sim U(0,1)$, $\xi_u^{\rm{req}}\sim U(0.8,0.9)$, and $\lambda_u \sim N(55,2.5^2)$. In addition, for text transmission users, we set $\varphi_u^{\rm{req}}\sim U(50,70)$ in ksuts/s and $\beta_u \sim N(0.2, 0.05^2)$. For image transmission users, we set $\varphi_u^{\rm{req}}\sim U(80,100)$ in ksuts/s and $\beta_u \sim N(0.1, 0.02^2)$. The threshold of the scores is set as $G_{\rm{th}}=0.5$. Further, we run the DeepSC model and DeepSC-VQA model to obtain the mapping relations $\xi_{q}^b=f_{\rm{Si}}(k_u,\gamma_{u}^b)$ and $\xi_{q}^b=f_{\rm{Bi}}(k_{ut}, k_{ui},\gamma_{ut}^b,\gamma_{ui}^b)$ based on the settings as $\mathcal{K}_{\rm{Si}}=\{1,2,\dots,20\}$, $\mathcal{K}_{\rm{Bi,t}}=\{2,4,6,8,10\}$, and $\mathcal{K}_{\rm{Bi,i}}=\{394,788,1576,2364,3152\}$, respectively.

We first verify the superiority of the developed QoE based formulation. Fig.~1 compares the QoE maximization and S-R maximization methods\cite{OurWork}. Here, the upper bound is obtained by assuming that the maximum number of users in each cell, i.e., ${\rm{max}}(|\mathcal{U}^b|,|\mathcal{M}|)$, can be served and their QoE can reach 1. Thus the upper bound is irrelevant to $G_{\rm{th}}$, which is compared to demonstrate the effectiveness of the proposed method. As $G_{\rm{th}}$ increases, the overall QoE of both methods decreases due to the tighter restriction. However, the overall QoE of the proposed method keep above 13 while the S-R maximization method shows worse performance, especially when $G_{\rm{th}}>0.5$. This implies that the proposed method can better adapt to the changes in the user requirements.

\begin{figure}
\vspace{-7pt}
  \centering
  \includegraphics[width=0.38\textwidth]{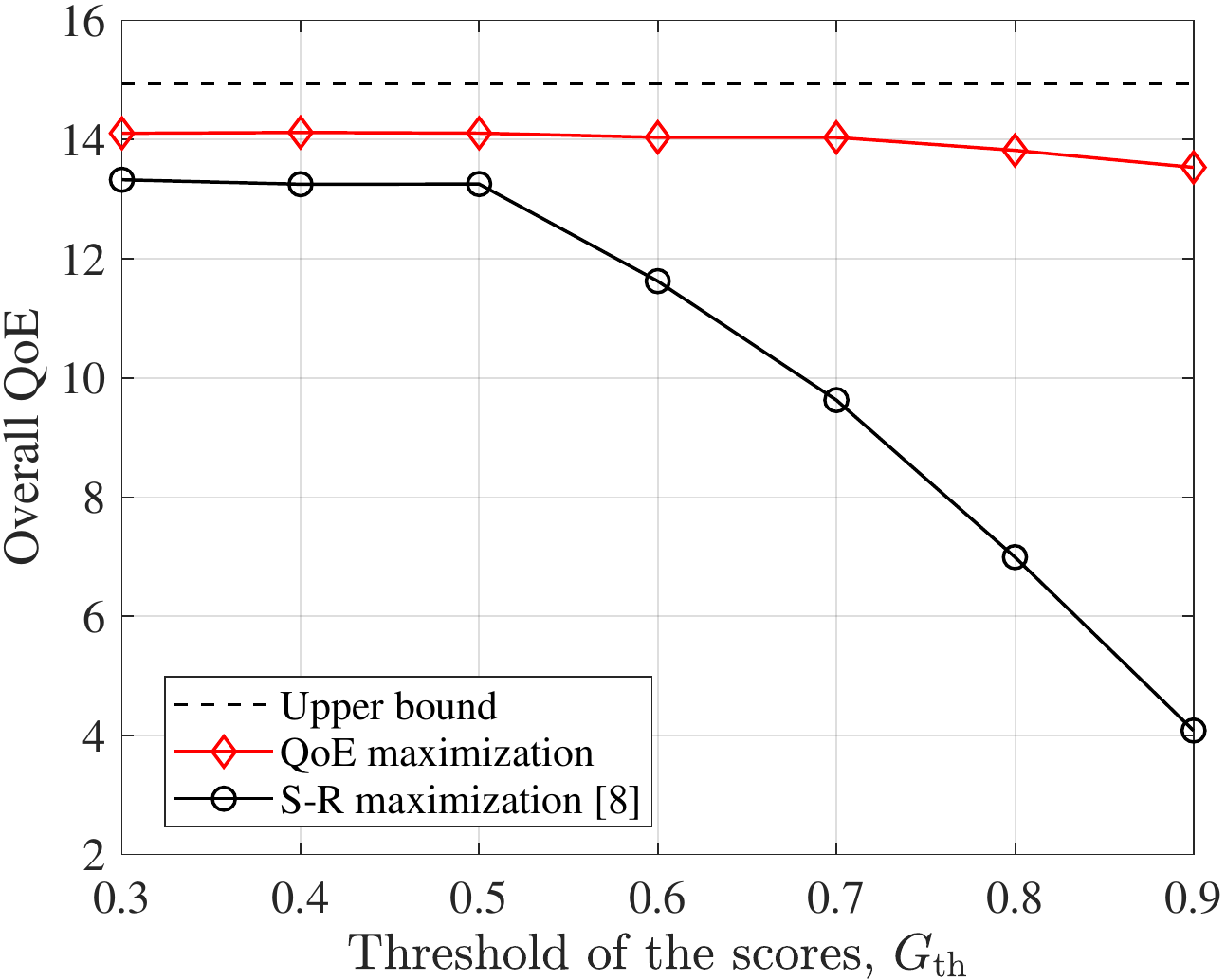}
  \vspace{-10pt}
  \caption{The overall QoE of the proposed QoE maximization method and the S-R maximization method \cite{OurWork} with $(N_{\rm{Si}},N_{\rm{Bi}})=(6, 6)$ and $|\mathcal{M}|=6$.}
\end{figure}

Fig.~2 illustrates the overall QoE versus the number of channels for different algorithms. Due to the unacceptable computation complexity of the exhaustive searching method as analysed in Section IV-B, we do not compare the proposed matching algorithm with that but with the upper bound of overall QoE, along with the random matching method. From this figure, the proposed matching algorithm outperforms the random method significantly and is very close to the upper bound. 
\begin{figure}
\vspace{-5pt}
  \centering
  \includegraphics[width=0.38\textwidth]{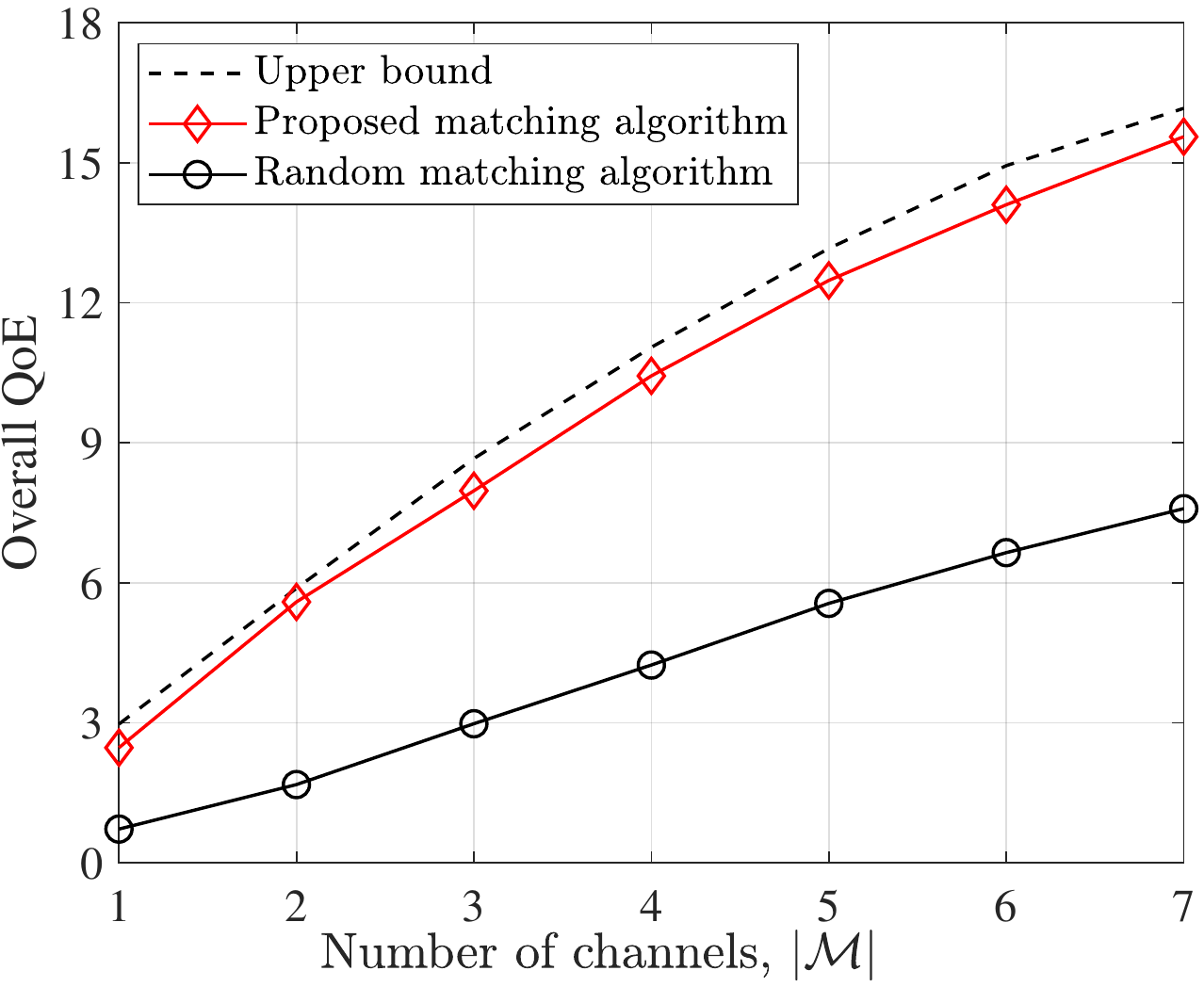}
  \vspace{-10pt}
  \caption{The overall QoE of different algorithms with $(N_{\rm{Si}},N_{\rm{Bi}})=(6, 6)$.}
\end{figure}



The proposed method with and without multi-cell cooperation are compared in Fig.~3. The method with multi-cell cooperation yields better performance than the one without multi-cell cooperation, which verifies the effectiveness of the proposed method in coping with the inter-cell interference.

\begin{figure}
\vspace{-5pt}
  \centering
  \includegraphics[width=0.408\textwidth]{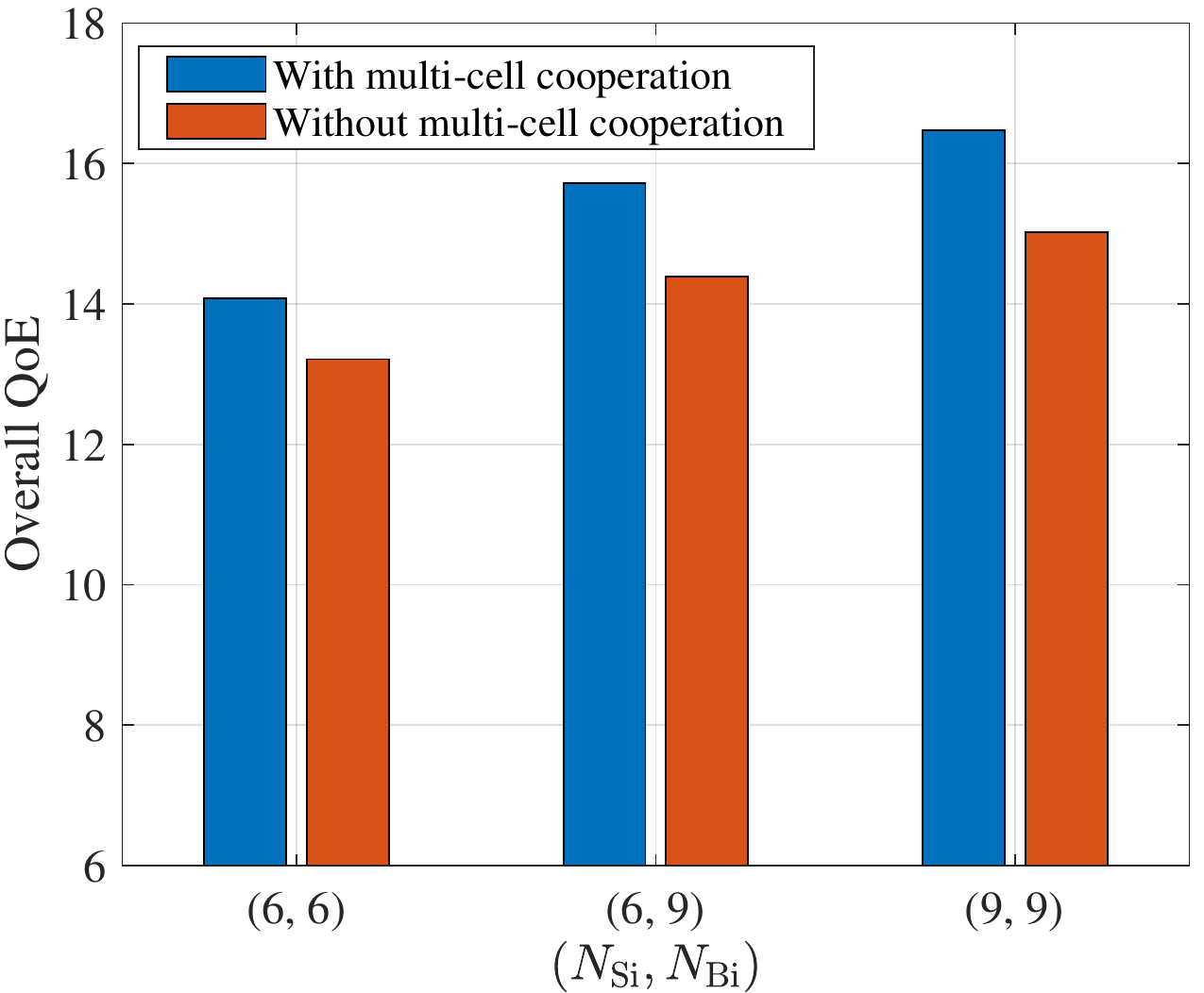}
  \vspace{-10pt}
  \caption{The overall QoE of the proposed method with and without multi-cell cooperation with $|\mathcal{M}|=6$.}
\end{figure}
\section{Conclusion}
In this paper, we studied the semantic-aware resource allocation in multi-task networks with multi-cell cooperation. Specifically, a novel quality of experience (QoE) model was developed for semantic communications in terms of semantic rate and semantic accuracy. Then we 
formulated a QoE maximization problem and solved it by a matching theory based solution.
Simulation results showed that the developed formulation can characterize the user requirements for semantic communications more properly and provide higher user satisfaction than the semantic transmission rate (S-R) based formulation. Moreover, the proposed algorithm outperforms all baselines significantly.
\vspace{-10pt}
\bibliographystyle{IEEEtran}
\bibliography{IEEEabrv}

\end{document}